\documentclass[cameraready]{Interspeech}

\title{Layer-wise Linear Probing of an Encoder-free Audio-capable Language Model}

\author[affiliation={1}, orcid=0000-0001-6779-5176]
{Laishram}{Rahul}

\author[affiliation={1}, orcid=0009-0008-4136-0236]
{Shruthi}{B.~S.}

\author[affiliation={2}, orcid=0000-0002-2239-3584]
{Parismita}{Gogoi}

\author[affiliation={3}, orcid=0000-0003-0743-9457]
{Sishir}{Kalita}

\author[affiliation={1}, orcid=0000-0003-2162-2776]
{Deepak}{KT}

\author[affiliation={1}, orcid=0000-0002-8135-7938]
{S R Mahadeva}{Prasanna}
\address{
    $^1$ Indian Institute of Information Technology Dharwad, India \\
    $^2$ DUIET, Dibrugarh University, Dibrugarh, India \\
    $^3$ Armsoftech.air, India
}
\email{\{rahul.24phddi03, shruthi.24mtrec01, deepak, prasanna\}@iiitdwd.ac.in,
parismita@dibru.ac.in, sisiitg@gmail.com}


\usepackage{comment}
\usepackage{seqsplit}

\begin{document}

\maketitle

\begin{abstract}
Encoder-free audio-capable multimodal language models provide a unique setting for studying
how speech representations emerge inside a language-model decoder. In
Gemma-4-12B, short waveform frames are projected directly into the shared
embedding space, and all subsequent contextual processing is performed by
the decoder layers. We use this architecture to examine where acoustic,
phonetic, segmental, and paralinguistic information becomes linearly
accessible across depth. Using the TIMIT corpus, we extract all 49
representations from the frozen model, including the embedding output and
the outputs of its 48 decoder layers. We train linear probes for
log-$F_0$, log-energy, voicing, phone identity, word-onset structure,
speaker identity, recorded speaker gender, and dialect-region labels. The
results show a broad progression in linear accessibility. Acoustic
properties are most strongly decodable in the early layers,
paralinguistic attributes are most accessible in the early-to-middle
layers, and phone identity and boundary-sensitive information reach their
highest performance around layer~11. Several properties remain strongly
decodable across broad near-peak regions rather than at a single uniquely
optimal layer. In deeper layers, phone identity and closed-set speaker
identity decline sharply, while voicing, log-$F_0$, and gender remain more accessible. These findings provide an initial map of how
speech information develops when the decoder is the first contextual stage
to process audio.
\end{abstract}


\section{Introduction}
Layer-wise probing has become a widely used approach for examining how
speech representations evolve across neural networks. In this methodology,
the pretrained network is frozen, a small \emph{linear probe} is attached
to each hidden layer, and the linear decodability of a property of interest
is measured across depth~\cite{alain2017probing,hewitt2019control}. Studies of self-supervised
speech encoders such as wav2vec~2.0~\cite{baevski2020wav2vec2},
HuBERT~\cite{hsu2021hubert}, and WavLM~\cite{chen2022wavlm} have often
reported a coarse representational progression: early layers retain local
acoustic information, intermediate layers make phonetic, prosodic, speaker,
and regional cues more accessible, and later layers become increasingly
oriented toward linguistic content or downstream objectives
~\cite{Yang2024Accent,
pasad2021layerwise,pasad2023comparative, Gogoi2025Tone,deLaFuente2024}.
Complementary analyses have shown that the distribution of speaker,
language, channel, and phonetic information also depends on model
architecture and training objective
~\cite{Chowdhury2024E2E,Ashihara2024CrossModel}. Related evidence further
suggests that representations commonly described as semantic may retain
substantial phonetic information~\cite{choi2024phonetic}.

Most audio-capable multimodal language models extend this architecture by
placing a speech encoder, acoustic tokenizer, or neural codec before the
language-model decoder. Examples include
SpeechGPT~\cite{zhang2023speechgpt},
AudioPaLM~\cite{rubenstein2023audiopalm},
Qwen-Audio and Qwen2-Audio
~\cite{chu2023qwenaudio,chu2024qwen2audio},
SALMONN~\cite{tang2024salmonn},
Qwen2.5-Omni~\cite{qwen25omni}, and
Moshi~\cite{moshi2024}. Although these systems differ substantially in
their objectives and audio interfaces, their decoders generally receive
representations that have already undergone contextual acoustic processing.
Consequently, probing only the language-model layers cannot determine
whether acoustic-to-phonetic transformations occurred in the decoder or in
the preceding audio front end.

Encoder-free audio language models create a different representational
setting. Rather than supplying the decoder with contextualized speech
features, they project short waveform or spectrogram segments directly into
the language-model embedding space. The decoder therefore becomes the first
contextual network responsible for transforming local acoustic observations
into representations that support phonetic and linguistic processing.
Mel-LLM, for example, projects Mel-spectrogram patches directly into an LLM
without a dedicated speech encoder
~\cite{fan2026llmreadspectrogram}.

Gemma-4-12B provides a suitable model for studying this setting. In
\texttt{google/gemma-4-12B}~\cite{gemma42026,gemma4hf}, a raw
$16$\,kHz waveform is divided into $640$-sample ($40$\,ms) frames, and a
linear projection maps each frame directly into the shared embedding space.
All subsequent contextual processing is performed by the model's 48
transformer decoder layers. This architecture allows us to examine where
acoustic, phonetic, segmental, and paralinguistic properties become linearly
accessible when no dedicated audio encoder precedes the decoder.

\subsection{Motivation and contributions}

We investigate where acoustic, phonetic, segmental, and paralinguistic
properties become linearly accessible across the decoder depth of an
encoder-free audio language model. Unlike conventional audio LLMs, Gemma~4
12B does not provide its decoder with representations produced by a
dedicated speech encoder. It therefore offers a direct setting for
examining whether a coarse acoustic-to-phonetic progression develops
within the language-model stack itself.

To the best of our knowledge, no prior work has mapped these properties
across a frozen encoder-free audio LLM using phone-aligned speech. We
present the resulting analysis as a descriptive map of representational
accessibility rather than as evidence that the probed properties are
causally used by the model. To reduce the possibility that probe
performance reflects only the probe or dataset, we include shuffled-label
controls at every layer~\cite{hewitt2019control}.

\noindent\textbf{Probing targets.}
We organize the analysis around a set of targets supported by TIMIT
annotations. The \emph{acoustic} group uses voicing, log-F0, and log-energy
as low-level controls. The \emph{phonetic} group probes fine phone identity,
with a clean-token analysis to account for $40$\,ms tokenization. The
\emph{segmental} group probes word onsets and an onset-versus-internal
transition control, asking whether word-level structure is distinguishable
from within-word phone changes. In parallel, we probe \emph{paralinguistic}
attributes, such as speaker identity, gender, and TIMIT dialect region, to track how
this information changes with depth.

\noindent\textbf{Contributions.}
This paper makes three contributions.
\begin{itemize}
\item We provide a layer-wise probing analysis of a frozen
encoder-free audio LLM, probing all $49$ representations of
\texttt{google/gemma-4-12B}~\cite{gemma4hf} on TIMIT without
fine-tuning the model.
\item We map acoustic, phonetic, segmental, and paralinguistic information
across depth using sample-aligned phone and word annotations, while
treating the model's $40$\,ms tokenization as an explicit source of label
blur.
\item We compare the resulting decodability profile with SSL speech
encoder hierarchies, using the comparison as a hypothesis about shared
speech-processing structure rather than as proof of causal mechanism.
\end{itemize}
The analysis is descriptive: it identifies where information is linearly
recoverable, while leaving causal tests of whether these properties affect
generation to future work.


\section{Experimental Setup}
\subsection{Model and representations}
We analyze \texttt{google/gemma-4-12B}, an encoder-free multimodal
language model that processes raw audio without a dedicated speech
encoder~\cite{gemma4hf}. The model divides the raw $16$\,kHz signal into
$640$-sample frames, corresponding to $40$\,ms per audio token. Each frame
is mapped
into the shared embedding space by a lightweight linear audio projection,
after which all contextual processing is performed by the model's $48$
transformer decoder layers. This design places much of the
acoustic-to-linguistic processing after the initial projection inside the
same layers that we can probe, making the model suitable for a layer-wise
decodability analysis.

For every utterance, we run one forward pass with hidden-state output
enabled and extract all $49$ representations: the embedding output
(layer~0) and the $48$ decoder layers (layers~1--48). Each retained vector
has dimension $3840$. Audio-token positions are identified using the
model's audio-token identifier, and only hidden states at those positions
are stored. Under the processor implementation used here, token $i$  corresponds to samples $[640i,640i+640)$, so an $N$-sample waveform should produce
$\lceil N/640\rceil$ audio tokens. Hidden-state extraction is performed in
bfloat16 without quantization.

We use \texttt{google/gemma-4-12B} at Hugging Face revision
\texttt{\seqsplit{56820d7d8cbe8e47975a53325439ed272e91cff2}}, with the
processor(feature extractor and tokenizer) loaded from the same revision, Transformers~5.10.2,
PyTorch~2.11.0, scikit-learn~1.9.0, and librosa~0.11.0 on a single
NVIDIA RTX PRO 4500 GPU; all seeds are fixed at~42.

\subsection{Input construction}
We probe the base model rather than an instruction-tuned conversational
variant. Accordingly, we use the base processor and provide only the audio
input, with no chat template and no textual instruction. A prompt such as
``Listen to this audio'' would remain in the model context and could
condition the hidden states at audio-token positions through causal
self-attention. Probing those states would then mix information from the
audio with information induced by the prompt. To avoid this confound, every
utterance is processed using the same prompt-free input construction.

Before the full extraction run, we perform an alignment sanity
check on the first utterance. The stored audio-token count must
exactly match $\lceil N/640 \rceil$, the model must return 49
hidden-state representations, and the hidden size must be 3840.
If the audio-token count does not match, extraction is aborted
before proceeding, because a token-count mismatch would shift
every frame-level label and invalidate alignment.


\subsection{Dataset}
All experiments use the training split of TIMIT~\cite{garofolo1993timit}. The
test split is not used in this work; instead, every task is evaluated with
the same five-fold cross-validation protocol on the training split, making the
layer-wise curves directly comparable across targets. The training split
contains $4{,}620$ utterances and yields $357{,}008$ audio tokens after
$40$\,ms framing.

Labels are derived from TIMIT's sample-level annotations: phone boundaries
from the \texttt{.PHN} files, word start and end samples from the
\texttt{.WRD} files, and verified speaker metadata for speaker identity,
gender, and dialect region. Unless otherwise stated, frame-level labels are
assigned by majority temporal overlap between the $40$\,ms token window and
the annotated segment.

For phone probing, we also compute a token-level purity score: the
fraction of the $640$-sample window covered by the majority-overlap phone.
Tokens with no overlapping phone, such as inter-word silence, receive no
purity value and are excluded from phone probing. A token is called
\emph{clean} if its purity is at least $0.80$. Under this definition,
$70.9\%$ of all audio tokens are clean, and $68.5\%$ of probe-eligible
non-silence tokens are clean. After excluding the specified silence-related labels, our probing inventory contains 57 phone classes. 

\subsection{Probing protocol}
We use GPU-accelerated PyTorch~\cite{paszke2019pytorch} linear probes for
computational efficiency. For each target and representation, we train a single
\texttt{nn.Linear} probe on frozen hidden states.  Classification probes
use cross-entropy loss, while regression probes use mean-squared error.
Weights are initialized from
$\mathcal{U}(-1/\sqrt{d},1/\sqrt{d})$, with $d=3840$, and biases are
initialized to zero. For each task and fold, seed~42 is reused across all
49 representations to ensure identical probe initialization.

Each probe is optimized for 1,000 full-batch Adam updates with learning
rate $0.01$, $\beta_1=0.9$, $\beta_2=0.999$, and
$\epsilon=10^{-8}$. Weight decay is set to $1/(CN)$ for classification
and $\alpha/N$ for regression, where $C=\alpha=1$ and $N$ is the number
of training examples in the fold. No class weighting is applied.

\noindent\textbf{Frozen splits.}
For each task, the token subsample and the cross-validation folds are
created once before iterating over layers. The same subsample and folds are
then reused at every layer and for the corresponding shuffled-label
control. This is essential for layer-wise comparison: differences between
layers should reflect changes in the representation, not changes in which
examples were sampled or how the train/test partitions were drawn.
Frame-level tasks are capped at $80{,}000$ tokens, sampled once with the
fixed seed.

\noindent\textbf{Cross-validation.}
Folds are speaker-disjoint wherever this is possible: no speaker appears in
both the training and test partitions of a fold. Classification tasks use a
stratified grouped five-fold split, which keeps speakers disjoint while
balancing target classes across folds. Regression tasks use grouped
five-fold splits with speaker as the grouping variable. The only exception
is speaker identification, where the label itself is the speaker; a
speaker-disjoint split would make this task undefined. For that task, we
use stratified five-fold cross-validation, so speaker identity should be
interpreted as closed-set decodability among known speakers rather than
generalization to unseen speakers.

\noindent\textbf{Standardization.}
Hidden-state features are standardized to zero mean and unit variance using
statistics computed only on the training partition of each fold; the same
transformation is then applied to the held-out partition. Regression
targets are standardized within each training fold as well. Negative held-out $R^2$ values are retained.

\noindent\textbf{Metrics.}
For classification, the primary metric is macro-averaged F1, with accuracy
reported as a secondary metric. Macro-F1 is computed against a fixed global
label set for each task, so folds remain comparable even when a rare class
is absent from a particular held-out fold. Speaker identification is the
one exception for reporting emphasis: because the task asks which speaker
produced the utterance, accuracy is treated as the headline metric and
macro-F1 is also reported. For regression, we report the coefficient of
determination, $R^2$. All reported values are means over the five folds,
and standard deviations across folds are used as the measure of dispersion.

\noindent\textbf{Baselines and controls.}
Every task includes a simple baseline computed under the same folds: a
majority-class predictor for classification and a predict-the-mean
predictor for regression. We also train a shuffled-label control probe for
every task. The control uses the same representations, folds, and training
procedure as the real probe, but with randomly permuted labels. It is run
at every layer rather than at a single layer, so evidence for a
target-specific signal requires the real curve to rise above its own
per-layer control curve.

\subsection{Probing tasks}

We evaluate nine prediction targets through ten probe settings, because phone identity is evaluated on both clean and all non-silence tokens.

Unless otherwise stated,
frame-level targets are evaluated on per-token hidden states, while
utterance-level targets are evaluated after mean-pooling all audio-token
representations within each utterance.
\begin{figure*}[t]
  \centering
  \includegraphics[width=0.98\textwidth]{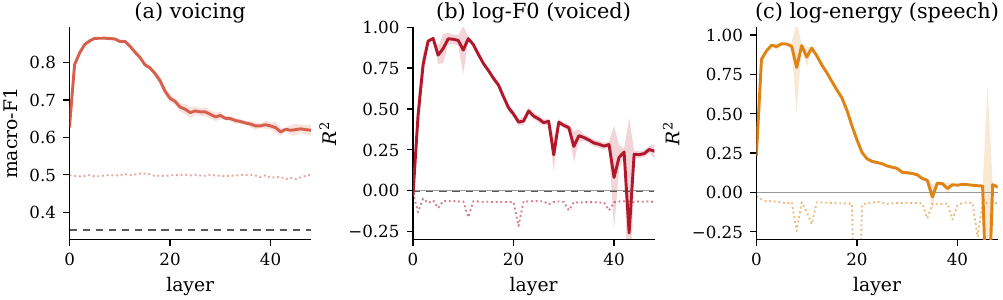}
  \caption{Layer-wise linear decodability of acoustic properties in
    \texttt{google/gemma-4-12B}: (a)~voicing (macro-F1), (b)~log-F0 on voiced
    frames ($R^2$), and (c)~log-energy on speech frames ($R^2$). Solid lines show
    five-fold cross-validated means, shaded bands show $\pm 1$ standard deviation
    across folds, and dotted lines show shuffled-label controls.}
  \label{fig:acoustic}
\end{figure*}

\noindent\textbf{Acoustic targets.}
The acoustic probes serve as low-level controls on token alignment and show
where basic signal properties are recoverable. We use three separate
targets rather than a single entangled acoustic probe. \emph{Voicing} is a
binary target indicating whether a frame is voiced, estimated with
pYIN~\cite{mauch2014pyin}.
\emph{log-F0} is a regression target defined only on voiced frames; unvoiced
frames are excluded rather than assigned a zero value, so the probe
measures pitch rather than a mixture of pitch and voicing. To align F0 with
token centers, pYIN is computed with a $320$-sample hop, so pYIN frame
$2i+1$ corresponds to the center sample $640i+320$ of token $i$.
\emph{log-energy} is a regression target on non-silence frames, computed as
the logarithm of the root-mean-square energy over the exact $640$-sample
token window. Because the model normalizes audio before projection,
absolute amplitude may be partly reduced; log-energy is therefore treated
as a weaker alignment check than voicing or F0.

\noindent\textbf{Phone targets.}
Fine phone identity is a $57$-class target assigned by majority overlap,
after excluding silence classes. To handle the coarse $40$\,ms tokenization
relative to phone durations, we run the phone probe twice. The primary
probe uses only clean tokens with purity at least $0.80$. A second probe
uses all non-silence tokens as a robustness check. The two runs are paired:
the clean-token folds are derived from the all-token folds, so the two
curves differ only in the purity filter.

\noindent\textbf{Segmental targets.}
The main segmental task is word-onset detection. A token is positive when a word onset falls within its 40 ms span and negative if it lies strictly inside a word; inter-word silence is excluded. Only word starts are labeled
positive, so this is an onset task rather than generic boundary detection.
The onset of the first word in an utterance is also excluded because the
initial silence-to-speech transition is a trivial cue. We additionally use
an onset-versus-internal control: positive examples are the same word-onset
frames, while negative examples are phone transitions that occur strictly
inside a single word. This control excludes utterance edges, word-final
boundaries, and speech-to-silence transitions. Strong performance on this separate contrastive task suggests that the probe is not relying exclusively on the presence of any phone transition.

\noindent\textbf{Paralinguistic targets.}
The paralinguistic probes are evaluated at the utterance level using
mean-pooled audio-token representations. We probe closed-set speaker
identity over the training speakers, the binary speaker-sex label
recorded in the TIMIT metadata, and the eight TIMIT dialect-region
labels. The dialect-region task measures how the linear accessibility
of broad regional speech variation changes across layers. Although
these targets do not form part of the acoustic-to-phone-to-segmental
sequence, they reveal how different utterance-level attributes are
distributed across decoder depth.

\subsection{Outputs}

For each task, we report the nominal peak layer, its mean score and
across-fold standard deviation, the baseline, and the margin over
baseline. To avoid overinterpreting a single maximum across 49
representations, we also report the earliest layer within one standard
error of the peak and the contiguous layer range achieving at least
95\% of peak performance. These are descriptive near-peak summaries,
not formal confidence intervals.
\begin{figure*}[t]
  \centering
  \includegraphics[width=0.98\textwidth]{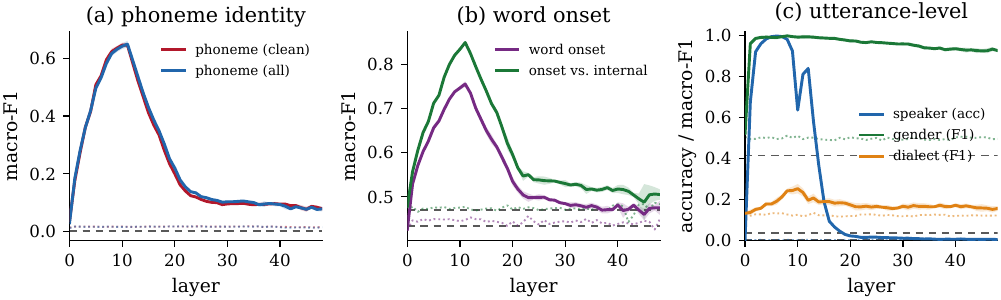}
  \caption{Layer-wise linear decodability of linguistic and paralinguistic
    targets. (a)~Phone identity (57 classes), comparing clean tokens with all
    non-silence tokens. (b)~Word-onset detection and the
    onset-versus-within-word-transition control. (c)~Utterance-level attributes
    from mean-pooled representations: closed-set speaker identity (accuracy),
    gender (macro-F1), and dialect region (macro-F1). Solid lines show
    five-fold cross-validated means, shaded bands show $\pm 1$ standard deviation
    across folds, dotted lines show shuffled-label controls, and dashed lines show
    chance or majority baselines.}
  \label{fig:linguistic}
\end{figure*}

\section{Results}
We first describe how each property changes with depth, then summarize
peak linear decodability by task. Throughout,
``peak layer'' denotes the layer at which a probe reaches its maximum under
the task's primary metric. All values are five-fold cross-validated means.
Layer indices run from $0$ (the embedding output) to $48$ (the final decoder
layer), and each probe is interpreted together with its baseline and
shuffled-label control.

\subsection{Layer-wise behavior of individual properties}
\noindent\textbf{Acoustic properties.}
Figure~\ref{fig:acoustic} shows the acoustic probe trajectories. The acoustic
targets become decodable early. log-F0, measured only on
voiced frames, rises from $R^2=-0.02$ at the embedding layer to
$R^2=0.93$ at layer~4. log-energy, measured on non-silence frames, peaks
slightly later with $R^2=0.94$ at layer~5. Voicing, which is a binary target indicating the presence of periodic excitation, peaks at layer~7 with macro-F1 $=0.87$. All three acoustic probes
are well above their shuffled-label controls, which remain near
$R^2=-0.06$ for the regression targets and macro-F1 $=0.50$ for voicing.
log-F0 reaches maximum linear decodability before the voicing classifier
reaches its maximum. This comparison should be read cautiously because log-F0 is evaluated only on already-voiced frames, whereas voicing is a
classification task over a different target population.

The acoustic curves also differ in how much information remains in deeper
layers. log-F0 remains partly decodable at layer~24 ($R^2=0.46$) and at
layer~48 ($R^2=0.24$), and voicing remains above chance at the final layer
(macro-F1 $=0.62$). log-energy decays more strongly, reaching only
$R^2=0.03$ by layer~48. We interpret this energy result cautiously because
the model normalizes audio before projection, which may reduce absolute
amplitude information independently of depth. The regression controls are
generally close to small negative $R^2$ values but exhibit several
high-variance negative excursions in deeper layers.
These excursions indicate fold or numerical sensitivity, so we emphasize
peak patterns supported by controls rather than isolated deep-layer
regression points.

\begin{figure}[t]
  \centering
  \includegraphics[width=0.95\columnwidth]{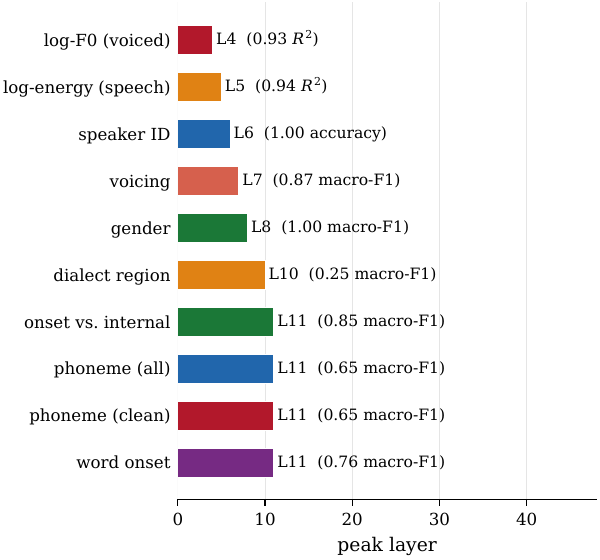}
  \caption{Peak linear decodability by task for encoder-free model. Each bar marks the layer at which a
    probe attains its maximum, ordered by peak layer; annotations give the peak
    value and metric.}
  \label{fig:peak-decodability}
\end{figure}

\noindent\textbf{Phone identity.}
Fine phone identity is strongly decodable, while the embedding
representation exposes only weak phone information compared with the
layer~11 peak. The clean-token phone probe peaks at layer~11
with macro-F1 $=0.646$ for $57$ phone classes, compared with a majority
baseline of $0.003$. At the embedding layer, macro-F1 is only $0.04$, close
to the shuffled-label control ($0.016$) and far below the mid-layer peak.
This indicates that phone identity becomes substantially more linearly
accessible by the middle layers.

Figure~\ref{fig:linguistic}(a) compares clean-token and all-token phone
probing. The clean-token and all-token phone probes are almost identical.
The all-token probe peaks at macro-F1 $=0.649$, also at layer~11. The
near-identical curves show that the layer-wise phone result is robust to
restricting evaluation to tokens with purity at least $0.80$. They do not
establish that boundary-straddling tokens carry information comparable to
clean tokens, because the all-token set contains many clean tokens and the
two evaluations overlap substantially. A stronger test would train a third
probe only on low-purity tokens, for example purity below $0.80$. After
the peak, phone decodability declines steadily, reaching macro-F1 $=0.13$
at layer~24 and $0.08$ at layer~48. The final layer remains slightly above
the shuffled-label control ($\approx0.015$), but most of the phonetic
signal is concentrated in the middle of the network.

\noindent\textbf{Segmental properties.}
Figure~\ref{fig:linguistic}(a,b) summarizes the phone and segmental probes.
Word-onset detection also peaks at layer~11, with macro-F1 $=0.76$ against
a majority baseline of $0.47$. This indicates that word-onset status becomes most linearly recoverable
within the same mid-layer region as phone identity.
However, word onsets often coincide with phone transitions, so a simple
onset detector could in principle be detecting acoustic change rather than
a word boundary.

We therefore evaluate a separate word-initial-versus-word-internal transition task. In this control, positive examples are word-onset frames and negative examples are phone
transitions that occur strictly inside a word. The task peaks at layer~11
with macro-F1 $=0.85$, above its majority baseline ($0.43$) and its
shuffled-label control ($0.48$). Because the negative class excludes utterance edges, word-final
boundaries, and speech-to-silence transitions, the result shows that the
representations distinguish word-initial from strictly within-word phone
transitions, beyond generic phone-transition detection. This is consistent
with boundary-sensitive information, although it does not by itself
establish an abstract lexical-boundary representation.

\noindent\textbf{Paralinguistic properties.}
Figure~\ref{fig:linguistic}(c) shows the utterance-level probes. Speaker
identity is highly decodable but sharply localized. It reaches
$99.7\%$ closed-set accuracy at layer~6, far above chance ($0.2\%$),
and then falls quickly: $1.6\%$ by layer~24 and approximately chance
($0.4\%$) by layer~48. Because this task uses stratified folds over
utterances from the same speaker inventory, it measures separability of
known training speakers rather than recognition of unseen speakers. Gender
peaks at layer~8 with macro-F1 $=0.998$ and remains high through the final
layer (macro-F1 $=0.93$). Dialect region is weaker but still above chance,
peaking at layer~10 with macro-F1 $=0.25$ against a majority baseline of
$0.035$. These trajectories suggest that closed-set speaker identity
becomes much less linearly accessible at depth, while a coarser attribute
such as gender remains strongly decodable.

\subsection{Peak linear decodability by task}

Figure~\ref{fig:peak-decodability} summarizes the nominal peak layer for
each probe, while Table~\ref{tab:peak_near_peak} reports the corresponding
near-peak ranges. The results indicate a broad progression in linear
accessibility across decoder depth. Acoustic properties peak in the early
layers, with log-$F_0$ at layer~4, log-energy at layer~5, and voicing at
layer~7. Utterance-level paralinguistic attributes peak in the
early-to-middle layers: closed-set speaker identity at layer~6, recorded
speaker sex at layer~8, and dialect-region classification at layer~10.
Phone identity, word-onset detection, and the onset-versus-internal
contrastive task all reach their nominal maxima at layer~11. This ordering
is qualitatively consistent with the coarse acoustic-to-phonetic
progression reported for dedicated SSL speech encoders, although the
architectures and layer indices are not directly comparable.

The nominal maxima should not be interpreted as uniquely optimal layers.
Because each peak is selected from 49 representations, small
cross-validation fluctuations may change the selected layer. The
near-peak summaries instead reveal partially overlapping representational
regions: speaker identity remains near peak across approximately
layers~3--8, voicing across layers~2--13, dialect-region information across
layers~8--10, and phone and boundary-sensitive tasks across approximately
layers~8--12. The apparent sequence
L4$\rightarrow$L5$\rightarrow$L6$\rightarrow$L7$\rightarrow$L8
$\rightarrow$L10$\rightarrow$L11 therefore represents a coarse ordering
of stages rather than statistically distinct transitions between
neighbouring layers.

Beyond the mid-layer region, the trajectories diverge. Closed-set speaker
identity and phone identity decline sharply, whereas voicing and log-$F_0$
remain moderately accessible and recorded speaker sex remains strongly
decodable through the final layer. log-energy approaches the
predict-the-mean baseline. These differences indicate selective changes in
linear accessibility rather than a uniform loss of speech-related
information with depth.
\begin{table*}[t]
\centering
\caption{Peak and near-peak probing performance across layers.}
\label{tab:peak_near_peak}
\small
\setlength{\tabcolsep}{4pt}
\begin{tabular}{lcccccc}
\toprule
Task & Metric & Peak layer: score $\pm$ SD & Baseline & $\Delta$ & Earliest 1-SE & Layers $\geq 95\%$ of peak \\
\midrule
phone (all) & Macro-F1 & L11: 0.6493 ± 0.0115 & 0.0023 & +0.6470 & L11 & L8–L11 \\
phone (clean) & Macro-F1 & L11: 0.6458 ± 0.0074 & 0.0026 & +0.6433 & L10 & L8–L11 \\
Word onset & Macro-F1 & L11: 0.7556 ± 0.0063 & 0.4698 & +0.2859 & L11 & L8–L12 \\
Onset vs. internal & Macro-F1 & L11: 0.8498 ± 0.0039 & 0.4334 & +0.4164 & L11 & L9–L12 \\
Voicing & Macro-F1 & L7: 0.8653 ± 0.0037 & 0.3524 & +0.5129 & L5 & L2–L13 \\
Log F0 & R² & L4: 0.9311 ± 0.0090 & -0.0025 & +0.9336 & L4 & L3–L4 \\
Log energy & R² & L5: 0.9444 ± 0.0021 & -0.0009 & +0.9452 & L5 & L2–L7 \\
Speaker ID & Accuracy & L6: 0.9970 ± 0.0008 & 0.0022 & +0.9948 & L6 & L3–L8 \\
Gender & Macro-F1 & L8: 0.9979 ± 0.0035 & 0.4137 & +0.5842 & L8 & L1–L33 \\
Dialect region & Macro-F1 & L10: 0.2520 ± 0.0230 & 0.0349 & +0.2171 & L9 & L8–L10 \\
\bottomrule
\end{tabular}
\vspace{1mm}
\begin{minipage}{0.98\textwidth}
\footnotesize
The earliest 1-SE layer is the earliest layer whose mean score is at least the peak mean minus the peak-layer standard error, with $\mathrm{SE}=\mathrm{SD}/\sqrt{5}$. The 95\% near-peak interval is the contiguous layer range containing the peak for which performance remains at least 95\% of the peak value.
\end{minipage}
\end{table*}

\section{Discussion}

The principal finding is that speech-related information is not uniformly
accessible across the decoder depth of Gemma-4-12B. Instead, the probes reveal
partially overlapping representational regions: local acoustic properties are
most strongly recoverable in the early layers, paralinguistic attributes are
prominent in the early-to-middle layers, and phone identity and
boundary-sensitive information become most accessible in a later region around
layers 8--12. The near-peak intervals further show that these properties often
form broad plateaus rather than uniquely optimal layers. The results therefore
support a coarse progression in representational accessibility, rather than a
strict layer-by-layer hierarchy. We relate the resulting decodability profile qualitatively to previously
reported SSL speech-encoder hierarchies, without treating their layer
indices or computational stages as directly equivalent.

The trajectories also differ substantially after their respective peaks.
Closed-set speaker identity and phone identity become much less linearly
accessible in deeper layers, whereas voicing, log-F0, and recorded speaker sex
remain comparatively well preserved. This pattern suggests that deeper
representations do not uniformly discard acoustic or paralinguistic
information. Instead, they selectively retain some broad attributes while
reducing the linear accessibility of finer-grained distinctions. The
near-perfect speaker-identification result should nevertheless be interpreted
carefully: because the probe is evaluated over the same closed set of speakers,
it measures separability among known TIMIT speakers rather than generalization
to unseen speakers or open-set speaker recognition.

The segmental probes provide evidence for sensitivity to word-initial
structure, but they do not establish that the model represents abstract lexical
boundaries. The word-initial-versus-word-internal transition task distinguishes
word-onset frames from phone transitions occurring strictly within words,
making it less likely that the classifier relies only on the presence of a
generic acoustic transition. However, the probe may still exploit correlated
cues such as word-initial phonotactics, stress, phone identity, duration,
prosody, or local silence. The result is therefore best interpreted as evidence
for boundary-sensitive information rather than a complete lexical-boundary
representation.

The TIMIT dialect-region label is the weakest of the evaluated targets.
Although its peak macro-F1 remains above the majority and shuffled-label
baselines, it is substantially lower than the scores obtained for the other
utterance-level attributes. This may reflect the broad and overlapping nature
of the TIMIT dialect regions, the limited duration of individual read-speech
utterances, or weak regional variation in the available material. We therefore
interpret the result as evidence that dialect-region information is only weakly
linearly accessible in the model representations.

\noindent\textbf{Limitations.}
The study is restricted to linear probes, a single clean read-speech corpus,
and five-fold cross-validation on the TIMIT training split. The reported
patterns may therefore differ for conversational or noisy speech, other
languages and dialects, cross-corpus evaluation, open-set speaker recognition,
or nonlinear probes. In addition, linear decodability indicates that a property
is accessible to a simple classifier, but it does not establish that the model
uses that property causally during generation. The log-energy analysis is also
affected by pre-projection audio normalization, and some deep-layer regression
controls exhibit optimizer-related variability. We therefore emphasize broad
near-peak regions and consistent separation from shuffled-label controls rather
than isolated maxima at individual layers.


\section{Conclusion}

We presented a layer-wise linear probing analysis of speech representations in the encoder-free Gemma-4-12B model. Using sample-aligned annotations from TIMIT, we examined acoustic,
phonetic, segmental, and paralinguistic information across all
49 representations: the embedding output and the outputs of the
48 decoder layers. The results support a broad representational progression rather than a sequence of uniquely optimal layers: acoustic properties are most accessible in the early layers, closed-set speaker identity and dialect-region information are strongest in the early-to-middle layers, and phone identity and boundary-sensitive targets occupy a later region around layers 8--12. The near-peak analysis further shows that several properties form extended plateaus, even when their nominal maxima occur at different layers. Across tasks, baseline-normalized probing trajectories indicate a common transition toward reduced linear decodability around layers 10--12, although the precise onset and rate of decline remain property-dependent.

The deeper layers exhibit selective rather than uniform loss of speech information. Phone identity and closed-set speaker identity decline substantially after their peaks, whereas voicing, log-F0, and gender remain more linearly accessible. These findings provide a descriptive map of how speech representations develop when the decoder is the first contextual stage to process audio. Because linear decodability does not establish causal use, future work should validate these patterns across additional corpora and speaking conditions and complement probing with intervention-based analyses.

The findings should be viewed as a case study of Gemma-4-12B rather than a general property of encoder-free audio LLMs, as comparable publicly available architectures at this scale remain limited.


\section{Generative AI Use Disclosure}
Generative AI tools were used during the preparation of this manuscript to assist with grammar correction, spelling checks, table generation and document formatting.

\bibliographystyle{IEEEtran}

\end{document}